\documentstyle[preprint,aps,12pt]{revtex}
\begin{document}

\preprint{\vbox{\hfill UTHEP-286\\ \null\hfill August, 1994}}

\title{ Gauge Fields  Emerging From Extra Dimensions \\
--- a Born-Oppenheimer approach ---}

\author{Tetsuo Hatsuda$^1$ and Hiroshi Kuratsuji$^2$}
\address{$^1$Institute of Physics, University of
Tsukuba, Tsukuba, Ibaraki 305, Japan}
\address{$^2$Department of Physics, Ritsumeikan University-BKC,
        Kusatsu 525-77, Japan}


\maketitle

\begin{abstract}

 We propose a dynamical mechanism to induce gauge fields
 in four dimensional space-time
  from a single scalar field or a spinor field in higher
 dimensions.  The Born-Oppenheimer treatment of the extra
 dimensions is an essential ingredient in our approach.
 A possible applications of the idea to
  low dimensional condensed matter systems
 and high temperature field theory  are also pointed out.
  This paper is an extended version of our previous unpublished work
 (SUNY-NTG-89-48, Jan. 1990).

\end{abstract}

\newpage

\noindent
{\bf  1. Introduction.}\ \ \
One of the ultimate goals in physics is to understand the fundamental
law of interactions acting between elementary particles.
Presently, it is generally believed
that the gauge theory, which is based on the principle of the
local gauge invariance, provides a fundamental form of
interactions.
 This strong
 requirement settles the form of interactions between matter and field
 in almost unique way.

However, it seems that there is still a room to look for
the origin of the gauge symmetry without imposing it from the outset.
A possible way is to extract the symmetry
a posteriori as a result of some simple dynamical mechanism.
 In what follows, we  shall develop a  theory along with this
thought. The basic idea has been already reported
 in our  unpublished/unsubmitted work \cite{HK}, and the
 present paper is its extended version.
 We will derive a massless gauge field only through the scalar
field or the Dirac field:
 The central point is that
  the gauge field emerges as an effective field as a result of
the smearing out the degrees of freedom associated with the ``extra
dimensions".  The idea may somewhat share the spirit with Kaluza-Klein
 theories \cite{KK}.  However,
 the gauge field is induced dymanically in our case, while
 it is already assumed from the outset in the Kaluza-Klein theories
 \cite{kiku}.
 The induced gauge field in our
  approach has formal analogy with that
  obtained by the Born-Oppenheimer approximation
 in  quantum mechanical
 systems such as the diatomic molecules \cite{SW}, which we will discuss
 more.

Let us start with a complex scalar field living in the
 (4+N)-dimensional space-time labeled by
 $ x^{M} = (x^{\mu}, \theta^i),$ $(\mu=0 \sim 3,\  i=5 \sim 4+N)$
 with the metric $g^{MM'}={\rm diag.}(1,-1, \cdot \cdot \cdot , -1)$.
 $x^{\mu}$ and $\theta^i$ denote the 4-dim. space-time coordinate and
 the extra N-dimensional space coordinate, respectively.
  Assume that the lagragian of the system has a simple form:
\begin{eqnarray}
\label{lagrangian}
  {\cal L}(x,\theta)  =  - \Phi^{\dagger} \partial_M \partial^M \Phi
                    -  [ c_1 \Phi^{\dagger} \Phi
              +  c_2 (\Phi^{\dagger} \Phi )^2 ],
\end{eqnarray}
and we will consider the case where $c_{1,2} \geq 0$ for simplicity.
 By using the standard Hubbard-Stratonovich transformation (the gaussian
 trick in the path integral) \cite{HS},
  \ (\ref{lagrangian}) is made equivalent with
\begin{eqnarray}
\label{meanfield}
{\cal L}(x,\theta) =               - \Phi^{\dagger}
             [\partial_{x}^2 +
             \partial_{\theta}^2 + c_1 +  \sigma(x,\theta)]
 \Phi  + {1 \over 4c_2}  \sigma(x,\theta)^2,
\end{eqnarray}
where $\sigma(x,\theta)$ is an independent
 auxiliary-field to be integrated out in the path integral
 together with $\Phi$.

 As far as the internal space is compactified,
 the following discussions do not depend on its
 topology nor the
 dimensinality N.  Therefore, we will take $N=1$
 and adopt a circle $S^1$ for the internal space
  in this section to demonstrate the essentail idea.
 In this choice,
  $\Phi(x,\theta)$ obeys the boundary condition
\begin{eqnarray}
\label{boundary}
  \Phi(x^{\mu}, \theta + 2\pi R) = \Phi(x^{\mu},\theta),
\end{eqnarray}
 where $R$ is the ``radius" of the internal space and
 has  a size of $O(1/($Plank mass)).
 We will assume that
$R$ is independent of  $x^{\mu}$ for simplicity unless otherwise
 is stated; this assumption is not essential for the following discussions.

 Let us adopt here the Born-Oppenheimer (BO) approach \cite{BO} and
 assume  that the fluctuation of $\Phi$ by the change of  $x^{\mu}$
 ($\theta$) is ``slow'' (``fast").  This corresponds to
 neglecting $\partial_x$ compared to $\partial_{\theta}$
 when one solves the dynamics in the internal space.
  This assumption is justified if the size of the internal space
 is small enough compared to the typical scale in the
 4-dim. world.  In the BO approximation,
 $ \Phi $ at each space-time point $x^{\mu}$ is
 expanded as
\begin{eqnarray}
\label{expansion}
  \Phi(x,\theta) = \sum_n \phi_n(x)f_n(\theta; x) .
\end{eqnarray}
Here $ f_n(\theta;x) $ is a complete set of functions
 in the internal space and is a solution of  the BO
 equation where $\partial_x$ is neglected
 by definition:
\begin{eqnarray}
\label{BO}
H_{_{BO}}  f_n(\theta;x) & = & \lambda_n (x) f_n (\theta;x)
\ \ \ {\rm with} \ \ \
 H_{_{BO}}  =  \partial_{\theta}^2 + c_1 +  \sigma (x,\theta),
\end{eqnarray}
where $\lambda_n(x)$ is an eigenvalue at each point $x^{\mu}$.
  Note that $ x$ and
$ \theta $ do not enter separately in $ f_n $, because
$\sigma$ is a function of both variables.
 $f_n$ satisfies the normalization and completness relations
\begin{eqnarray}
\label{normalization}
  \int  d\theta \  f_n^{*}(\theta;x) f_m(\theta;x) = \delta_{nm},\ \ \ \
    \sum_n f_n^{*}(\theta;x) f_m (\theta^{'};x) =
                        \delta(\theta-\theta^{'}) .
\end{eqnarray}

 The effective Lagrangian $\hat{\cal L}$
 in the ordinary space-time is defined as
\begin{eqnarray}
\label{effective}
   \hat{\cal L}(x)  = \int d\theta \ L(x,\theta) \ \ .
\end{eqnarray}
 Substuting the decomposition (\ref{expansion}) into
 (\ref{lagrangian}) and using (\ref{normalization}),
 one  obtains
\begin{eqnarray}
\label{bose1}
  \hat{\cal L} = [(\partial^{\mu}-iA^{\mu}_{ln})\phi_n]^{\dagger}
               (\partial_{\mu} - i A_{\mu}^{lm})\phi_m
              - \int d\theta \ [\lambda_n \phi^{\dagger}_n \phi_n
    - \ {1 \over 4c_2} \sigma^2 (x,\theta)],
\end{eqnarray}
with
\begin{eqnarray}
\label{gaugefield}
   A^{\mu}_{nm}(x) = i  \int d\theta \ f_n^{*} (\theta;x)
 \partial^{\mu}f_m (\theta;x)  ,
\end{eqnarray}
and  $\partial^{\mu} = \partial / \partial x_{\mu}$.
Because of the decomposion (\ref{expansion}), $\hat{\cal L}$
 in (\ref{bose1}) has an
 obvious local U(1) gauge invariance, namely,
\begin{eqnarray}
\label{gaugesymmetry}
    \phi_n  \rightarrow \phi_n e^{i\alpha},\ \ \ \ \
         f_n \rightarrow f_n e^{-i\alpha} \ \ \   {\rm or}
   \ \ \        A^{\mu}_{nm} \rightarrow A^{\mu}_{nm}
                   + \delta_{nm} \partial^{\mu}\alpha.
\end{eqnarray}
 Note here that $\sigma$ is gauge-singlet
 by definition. From the above  transformation property,
 $ A_{\mu}^{nm} $ can be identified with a U(1) gauge field induced by
  the
 existence of the internal space.

 There are several ways to the non-abelian
  generalization of the above result. Let us present two possible
 examples \cite{foot1}. (i)
  If there exists $k$-fold degeneracy for particular mode $n$ in the
 BO equation (\ref{BO}), the rotation among the degenerate modes
 with the same eigenvalue $\lambda_n$
 induces $U(k)$ gauge symmetry and associated $U(k)$ Yang-Mills (YM)
 field $A_{\mu}^{ab}(x)$ with $a,b = 1 \sim k$ ($n,m$ indices are
 suppressed here.)
 This is easily seen by the decomposition with an explicit
 label $a$; $\Phi =  f^{(a)}(\theta;x) \phi^{(a)}(x)$ ($a=1 \sim k$ and
 the label
 $n$ is not shown explicitely here).
 The local gauge invariance reads
 $\phi \rightarrow U \phi, f \rightarrow f U^{\dagger}$ where
 $U$ acts on the suffix $a$.
 (ii) Suppose we
   introduce a  scalar field
 $\Phi^{ab}(x,\theta)$ ($a,b=1 \sim k$) transforming as
 $U_{global} \Phi U_{global}^{\dagger}$ under the $U(k)$
 global rotation.
 Then, the BO decomposition for the $k \times k$ matrix
 $ \Phi^{ab} = \sum_n \sum_c  f_n^{ac}(\theta;x) \phi_n^{cb}(x)$
  naturally induces a local $U(k)$ gauge invariance
  $\phi \rightarrow  U_{local} \phi,  f_n \rightarrow f_n U_{local}^{\dagger}$
 where $U_{local}$ acts on the suffix $c$.
 Associated $U(k)$ YM field  transforms
 $A_{\mu} \rightarrow
U_{local} (A_{\mu} + i \partial_{\mu} )U_{local}^{\dagger}$.
  In the latter example, one extra-dimension is enough to
 accomodate the YM field, which is in contrast to the Kaluza-Klein theories
  where higher extra-dimensions are neccesary\cite{KK}.

Among a series of the scalar fields $\phi_n(x)$, the one corresponding to the
 lowest eigenvalue $\lambda_0$ will survive at low energies.
  Thus, retaining only $n=m=0$,  one arrives at
a simply effective lagrangian
\begin{eqnarray}
\label{bose2}
  \hat{\cal L} = (D_{\mu}\phi)^{\dagger}
               (D^{\mu} \phi)
              -  \int d\theta [\lambda \phi^{\dagger} \phi - {1 \over 4c_2}
  \sigma^2],
\end{eqnarray}
where $D_{\mu} \equiv \partial_{\mu} - i A_{\mu}$ and we have suppressed the
 label $n=m=0$.  The functional integral over $\sigma$ should  still
 be carried out to get a true low energy action, which will
 be discussed in the next section.

 A comparison of our approach with the BO treatment of the
 diatoms is worth mentioning here.
  Our field variables  $f_n$ and $\phi$  correspond to
  the electronic wave function and the nuclear wave function
  for the diatomic molecule, respectively.
   In the BO approximation of the diatom, the electronic wave function is
 solved as if the nuclear coordinate ${\bf X}$ is an external
 parameter, which induces the ``gauge field"
 in the effective hamiltonian for the nuclear motion \cite{SW}.
 This is quite analogous to our BO treatment of the internal space.
 A slight difference is that the object we are concerned with is
the field variables, while it is the wave functions
in the case of diatoms.

\vspace{0.8cm}

\noindent
{\bf 3. Non-triviality of the induced gauge field.}\ \ \
Up to now, we have neither given an explicit form of
 $f_n(\theta;x)$ nor discussed whether the induced $A_{\mu}$
 has nontrivial field strength $F_{\mu \nu}$.
 To answer these questions, let us go back to the BO equation
 (\ref{BO}).  If $\sigma$ is zero (i.e. $c_2 =0$),
 $f_n$ is readily solved as
\begin{eqnarray}
\label{freecase}
f_n^{(0)} (\theta;x) = {1 \over \sqrt{2 \pi R}} e^{in/R}
\ \ \ {\rm with} \ \ \ \lambda_n^{(0)}
 = c_1 +  ({n \over R})^2.
\end{eqnarray}
 From this, one observes the following facts:
 (i) If $R$  is not a function of $x$, the induced gauge field
 $A_{\mu}$ is trivialy zero and so does the field strength
 $F_{\mu \nu}$.
(ii) Even if $R$ is assumed to be $x$ dependent quantity,
 $F_{\mu \nu}$ is still zero since $A_{\mu} $ can be written
 as a total divergence.

When $c_2 \neq 0$, the gauge field becomes non-trivial
 even if $R$ is $x$ independent.
 This can be seen explicitely by calculating
  $f_n$ in the first order perturbation
 with respect to $\sigma$;
\begin{eqnarray}
\label{firstorder}
f_n(\theta;x) \simeq  \mid n \rangle +
 \sum_{l \neq n} { \mid l \rangle \langle l \mid  \sigma \mid n \rangle
 \over \lambda_n^{(0)} - \lambda_l^{(0)}  },
\end{eqnarray}
where we have used  abbriviated notations
 such as $f_n^{(0)} = \mid n \rangle$ and
 $  \langle l \mid   \sigma \mid n \rangle
 = \int d\theta \ f_l^{(0)*}  \sigma(x,\theta) f_n^{(0)} $.
 The induced gauge field corresponding to (\ref{firstorder}) becomes
 e.g.
\begin{eqnarray}
\label{firstordergauge}
A_{\mu}^{nn} (x) = - i \sum_l
 { \langle n \mid \sigma \mid l  \rangle
 \langle l \mid \partial_{\mu} \sigma \mid n \rangle
 \over (\lambda_n^{(0)} - \lambda_l^{(0)} )^2 },
\end{eqnarray}
which produces $F_{\mu \nu} \neq 0$.

 So far, $A_{\mu}$ is given as an implicit function of the
 ``background'' field $\sigma(x,\theta)$.
 To treat the induced gauge field as a dynamical one, we have to make
 a functional  integration over $\sigma$.
 For this purpose, let us insert an unity
$1 = \int  [dB_{\mu} (x)] \ \delta (B_{\mu}- A_{\mu})$
 into the partition function with the lagrangian (\ref{bose2}),
  which results in
\begin{eqnarray}
\label{semifinalpart}
Z = \int [dB_{\mu}][d\phi d\phi^*] e^{i \int d^4x \ \mid D_{\mu}\phi \mid^2}
 \int [d\sigma] \delta(B_{\mu}-A_{\mu})
e^{i \int d^4x d\theta \ (-\lambda \phi^{\dagger}\phi
 + {1 \over 4c_2} \sigma^2 )}\ \ \ .
\end{eqnarray}
Here $D_{\mu}$ is redefined as $ \partial_{\mu} - i B_{\mu}$.
 (\ref{semifinalpart}) has a gauge symmetry
 with simultaneous changes
 $\phi \rightarrow \exp[i \alpha] \phi,
 A_{\mu} \rightarrow A_{\mu} + \partial_{\mu} \alpha$ and
 $B_{\mu} \rightarrow B_{\mu} + \partial_{\mu} \alpha$.
 After the $\sigma$-integration, $A_{\mu}$ disappears from $Z$, thus
  the gauge symmetry carried  by
 $\phi$ and $B_{\mu}$ remains as  a remnant of the
  original symmetry carried by $\phi$ and $A_{\mu}$ in (\ref{gaugesymmetry}).
 This implies that only the
   gauge invariant terms composed of $\phi$ and $B_{\mu}$ appears
 in $Z$ after  the $\sigma$-integration, i.e.
\begin{eqnarray}
\label{finalpart}
Z = \int [dB_{\mu}][d\phi d\phi^*] e^{i \int d^4x \
(\mid D_{\mu}\phi \mid^2 - c_3 \phi^{\dagger}\phi
 - c_4 F_{\mu \nu}^2 + \cdot \cdot \cdot )},
\end{eqnarray}
with constants $c_3, c_4 \cdot \cdot \cdot$
 and $F_{\mu \nu} \equiv \partial_{\mu} B_{\nu} - \partial_{\nu} B_{\mu}$.
The exponent in (\ref{finalpart}) is the final low energy effective
 action where  $\phi$ and $B_{\mu}$ are the independent
 fields. The kinetic term of $B_{\mu}$ is induced by the
 quantum fluctuation of $\sigma \sim \Phi^{\dagger} \Phi$.
 The situation is similar to
 that in the CP$^n$ model \cite{CPN}
where the kinetic term of the composite gauge field
 is induced dynamically.  The analogy can be seen more closely
 by using  $\delta(B_{\mu} - A_{\mu}) =
 \int [dC_{\mu}] \exp (i C^{\mu}(B_{\mu} - A_{\mu}))$
 in (\ref{semifinalpart}).
 A major difference from the CP$^n$ model is that
 our composite gauge field $A_{\mu}$ is quite an implicit function
 of $\sigma$, while the gauge field in the CP$^n$ model is
 simply proportional to $z^* \partial_{\mu} z$ with $z$ being
 the original CP$^n$ field.  This make the
 evaluation of the coeeficients $c_3, c_4, \cdot \cdot \cdot$
 more involved and we will not pursue it here except for the
 general consideration.

\vspace{0.8cm}

\noindent
{\bf 4. Gauge field induced by a spinor.}\ \ \
We now turn to the gauge field induced by a Dirac field.
 Since most of the manipulations are the same with the boson case,
 we will just show the outline of the derivation here.
 Let's start with a massless Dirac field
 in 4+N dimensions with vector-type self interaction
 as an example;
\begin{eqnarray}
{\cal L} = \bar{\Psi} i \Gamma^M \partial_M \Psi
 - c_2 (\bar{\Psi} \Gamma_M \Psi)^2 ,
\end{eqnarray}
where $\Gamma_M$ is the 4+N dimensional gamma matrices satisfying
$ \{ \Gamma^M,\Gamma^{M'} \} = 2g^{MM'}$.
   After the Hubbard-Stratonovich transformation,
 the equivalent lagrangian reads,
\begin{eqnarray}
{\cal L} = \bar{\Psi} i \Gamma^M (\partial_M - i \omega_M) \Psi
 - {1 \over 4c_2} \omega^2_M(x,\theta) .
\end{eqnarray}
$\omega_{\mu}$ here is an auxiliary field and has nothing
 to do with the induced gauge field we are looking for.

Since $\tilde{\Gamma}^M \equiv  i\Gamma^0\Gamma^1\Gamma^2\Gamma^3
 \cdot
 \Gamma^M$ has a property
$ \{ \tilde\Gamma^{\mu},\tilde\Gamma^{\nu}\} = 4g^{\mu\nu},
 \{ \tilde\Gamma^i,\tilde\Gamma^j \} = 4g^{ij},
 [\tilde\Gamma^{\mu},\tilde\Gamma^i]= 0$
 with $\mu,\nu = 0 \sim 3$ and $i,j = 5 \sim 4+N$,
 one can choose a product representation as usual,
  $\tilde{\Gamma}_{\mu} = \gamma_{\mu}^{(4)} \otimes I^{(N)}$
 and $\tilde{\Gamma}_{i} = I^{(4)} \otimes \gamma_i^{(N)}$
 where $\gamma^{(L)}$ and $I^{(L)}$
  are the Diral matrices and unit matrix in $L$-dimensions, respectively.
 The Born-Oppenheimer equation corresponding to
 (\ref{BO}) reads
\begin{eqnarray}
i\gamma^{i}(\partial_{i}-i\omega_{i})f_n(x_{\mu},\theta)
= \lambda_n f_n(x_{\mu},\theta),
\end{eqnarray}
with $\Psi = \psi_n(x) \otimes f_n(\theta;x)$, and the
 effective lagrangian is written as
\begin{eqnarray}
\hat{\cal L}  =   \bar{\psi}_n
 i \gamma^{\mu} [\delta_{nm} (\partial_{\mu} -i \omega_{\mu})
 - i A_{\mu}^{nm}] \psi_m - \int d\theta \
 (\lambda_n \bar{\psi} \psi - \omega_{\mu}^2),
\end{eqnarray}
with
\begin{eqnarray}
A_{\mu}^{nm}(x)  = i \int d\theta f_n^{\dagger}\partial_{\mu}f_m .
\end{eqnarray}
The local gauge invariance $\psi_n \rightarrow \exp(i \alpha)\  \psi_n$,
 $A_{\mu}^{nm} \rightarrow A_{\mu}^{nm} + \delta_{nm} \partial_{\mu} \alpha$
is obvious in the above formula.
 Final low energy effective lagrangian ${\cal L}_{eff}$ is obtained
 by taking $n=m=0$ and integrating out $\omega_{\mu}$;
\begin{eqnarray}
{\cal L}_{eff} & = &  \bar{\psi}
 \gamma^{\mu} i \gamma_{\mu} D^{\mu}
  \psi - c_3  \bar{\psi} \psi - c_4 F_{\mu \nu}^2 + \cdot \cdot \cdot,
\end{eqnarray}
with $D_{\mu} \equiv \partial_{\mu} -i B_{\mu}$ and
 $F_{\mu \nu} = \partial_{\mu}B_{\nu} - \partial_{\nu} B_{\mu}$.

\vspace{1.2cm}

\noindent
{\bf 5. Further directions.}

\noindent
(I) We have so far concentrated on the vector type gauge fields.
 However, by generalizing the
 local field $\Psi(x,\theta)$ to the string field
 $\Psi(x(\sigma),\theta)$ or the membrane field
 $\Psi(x(\sigma_1,\sigma_2),\theta)$, one may
 induce antisymmetric tensor fields such as
 $A_{\mu \nu}$ and $A_{\mu \nu \lambda}$ which are the
 gauge fields associated with the motion of the string
 or the membrane as is discussed in ref.\cite{nambu}.

\vspace{0.3cm}

\noindent
(II) Field theories at high temperature $(T)$ is a possible place
 to apply the idea in this paper.
  In fact, in the imaginary-time formulation of the
 finite $T$ field theories,  the size of the compactified ``time''
 direction is $1/T$.
  Thus, at high $T$,  emergence
 of an induced gauge field is expected after eliminating  the
 modes in the time-direction through the BO procedure.
 This may provide us with a new method
 to construct high $T$ effective field theories \cite{finiteT}.

\vspace{0.3cm}

\noindent
(III) Application of our idea to the low-dimensional condensed matter
 system is also interesting.  Instead of introducing  the
 hypothetical extra dimensions, let us imagine to curl up
 a  two dimensional sheet (on which electrons are living) to a tube.
 If the diameter of the tube
 is small enough, the system can be treated
 almost like a one-dimensional system ``except'' for the
 gauge field induced by the curling-up procedure.
 This gauge field is a remnant of the wave functions of the electrons
 moving  in the compactified
 dimension, and it  will affect
 the  dynamics of the electrons moving along the
 uncompactified direction.
 The carbon nanometer tube, which was recently
 discoverd in the laboratory \cite{NEC},
  may be used to study such a novel phenomenon.
 Further works along these lines  line are  under investigation
  and will be reported elesewhere \cite{HKC}.

\vspace{1.2cm}

\noindent
{\bf 6. Summary.}\ \ \
In summary, we proposed a dynamical mechanism to induce the spin-one gauge
 fields from the scalar or spinor fields in higher dimensions.
 The Born-Oppenheimer
 treatment for  the compact internal space naturally
 induces gauge fields, which is analogous to the similar
 phenomena in quantum
 mechanical systems.  The idea could possibly
 be applied also to the low dimensional electron systems and the
 high $T$ field theories.

\vspace{3cm}

\noindent
{\bf Acknowledgements.} \ \ \
 We thank M. Hatsuda for the discussion on the
 antisymmetric tensor fields and K. Yabana for informing us the
 references on carbon nanometer tube.

{\bf}


\begin{thebibliography}{10}

\bibitem{HK} T. Hatsuda and H. Kuratsuji,
``Gauge fields emerging from extra spaces",
Stony Brook  preprint SUNY-NTG-89-48 (Jan. 1990) unpublished.


\bibitem{KK} ``Modern Kaluza-Klein Theories", Frontiers in Physics, eds.
 T.Appelquist and A.Chodos, ( Addison-Wesley, Menlo Park, 1987).

\bibitem{kiku}
  There is also a recent work
 having similar motivation with ours
  (K. Kikkawa, Phys. Lett. {\bf B297} (1992) 89.)
   However,  the gauge field  in higher dimensions
 is explicitely introduced in that work, which is quitely
different from our theory.

\bibitem{SW} ``Geometric Phases in Physics", eds.
 A.Shapere and F.Wilzcek, (World Scietific, Singapore, 1989).


\bibitem{HS} R. L. Stratonovich, Dokl. Akad. Nauk S.S.S.R. {\bf 115}
 (1957) 1907 (Sov. Phys. Dokl. {\bf 2} (1958) 416);
 J. Hubbard, Phys. Rev. Lett. {\bf 3} (1959) 77.


\bibitem{BO} M. Born and J. Oppenheimer, Ann. Phys. (Leipzig) {\bf 84}
 (1927) 457.



\bibitem{foot1}
 There is a non-abelian symmetry U($\infty$) inherent in
 the first term of (\ref{bose1}), namely
$ \phi_n \rightarrow (U\phi)_n,\
      f_n \rightarrow (fU^{\dagger})_n,
   A^{\mu} \rightarrow U (A^{\mu}+i\partial^{\mu})
                   U^{\dagger}.$
 This symmetry is,  however, badly broken by the second term in
 (\ref{bose1}) and we will not consider it here.



\bibitem{CPN} A. D'Adda, M. Luscher and P. DiVecchia, Nucl. Phys.
 {\bf B146} (1978) 63.

\bibitem{nambu} Y. Nambu, Phys. Rep. {\bf 23C} (1976) 250.



\bibitem{finiteT} For the recent developement of the
 effective field theory at high $T$
 and the dimensional reduction, see the recent review
 P. Lacok and T. Reisz, Nucl. Phys.  B (Proc. Suppl.)
 {\bf 30} (1993) 307 and the references therein.


\bibitem{NEC}  S. Iijima, Nature, {\bf 354} (1991) 56;\\
 T. W. Ebbesen and P. M. Ajayan, Nature, {\bf 358} (1992) 220.

\bibitem{HKC} T. Hatsuda and H. Kuratsuji, under investigation.


\end{thebibliography}
\end{document}